\documentclass[journal=aelccp]{achemso}

\usepackage{chemformula}
\usepackage[T1]{fontenc}

\usepackage[colorlinks=true,linkcolor=blue,citecolor=blue,urlcolor=blue]{hyperref}

\makeatletter
\@ifpackageloaded{hyperref}
  {\newcommand{\hrefdoi}[1]{, \href{http://dx.doi.org/#1}{DOI: #1}}}
  {\newcommand{\hrefdoi}[1]{}}
\@ifpackageloaded{hyperref}
  {\newcommand{\hrefarxiv}[2]{, \href{http://arxiv.org/abs/#1}{arXiv:#1 [#2]}}}
  {\newcommand{\hrefarxiv}[2]{}}
\makeatother

\hypersetup{
    colorlinks=false,
}

\newcommand*{\citen}[1]{
  \begingroup
    \romannumeral-`\x 
    \setcitestyle{numbers}
    \cite{#1}
  \endgroup   
}

\usepackage{url}
\usepackage{gensymb}
\usepackage[nameinlink,noabbrev,capitalize]{cleveref}
\usepackage{multirow}

\author{Aarti Tiwari}
  \affiliation{Department of Physics, Technical University of Denmark (DTU), Fysikvej 311, 2800 Kgs. Lyngby, Denmark}
  \altaffiliation{These authors contributed equally to this work}
\author{Hendrik H. Heenen}\altaffiliation{These authors contributed equally to this work}
  \affiliation{Department of Physics, Technical University of Denmark (DTU), Fysikvej 311, 2800 Kgs. Lyngby, Denmark}
  \altaffiliation{These authors contributed equally to this work}
\author{Anton Simon Bj\o rnlund}\altaffiliation{These authors contributed equally to this work}
  \affiliation{Department of Physics, Technical University of Denmark (DTU), Fysikvej 311, 2800 Kgs. Lyngby, Denmark}
  \altaffiliation{These authors contributed equally to this work}
\author{Degenhart Hochfilzer}
  \affiliation{Department of Physics, Technical University of Denmark (DTU), Fysikvej 311, 2800 Kgs. Lyngby, Denmark}
\author{Karen Chan}
  \affiliation{Department of Physics, Technical University of Denmark (DTU), Fysikvej 311, 2800 Kgs. Lyngby, Denmark}
  \email{kchan@fysik.dtu.dk}
\author{Sebastian Horch}
  \affiliation{Department of Physics, Technical University of Denmark (DTU), Fysikvej 311, 2800 Kgs. Lyngby, Denmark}
  \alsoaffiliation{Center for Bachelor of Engineering Studies, Technical University of Denmark (DTU), Lautrupvang 15, 2750 Ballerup, Denmark}
  \email{shor@dtu.dk}

\title
  {Electrochemical oxidation of CO on Cu single crystals under alkaline conditions}

\abbreviations{Cu, CV, SC}
\keywords{American Chemical Society, \LaTeX}

\begin{document}

\begin{abstract}
    We perform a joint experimental-theoretical study of the electrochemical oxidation of CO on copper (Cu) under alkaline conditions. Using cyclic voltammetry on Cu single crystal surfaces, we demonstrate that both Cu terraces and steps show CO oxidation activity at potentials just slightly positive (0.03--0.14~V) of the thermodynamic equilibrium potential. The overpotentials are 0.23--0.12~V lower than that of gold ($\approx 0.26$~V), which up until now has been considered to be the most active catalyst for this process. Our theoretical calculations suggest that Cu's activity arises from the advantageous combination of simultaneous *OH adsorption under CO oxidation potentials and surmountable *CO-*OH coupling barriers. Experimentally observed onset potentials are in agreement with the computed onsets of *OH adsorption.  We furthermore show that the onsets of *OH adsorption on steps are more affected by *CO-*OH interactions than on terraces due to a stronger competitive adsorption. Overall, Cu(100) shows the lowest overpotential (0.03~V) of the facets considered.  
\end{abstract}

In this paper, we revisit the electrochemical (EC) interaction of copper (Cu) with CO at oxidative potentials under alkaline conditions. The expected reaction, the EC oxidation of CO on Cu seems to have been overlooked since the early works of Kita \emph{et al.\@} in the 1980s, probably because they concluded that Cu was inactive for CO oxidation \cite{Kita1985}. This conclusion resulted in most subsequent research being focused on Au, leading to the understanding that it is the most active catalyst for EC CO oxidation \cite{Kita1985, Chang1991, Edens1996, Blizanac2004a, Rodriguez2010}. When we tried to reproduce Kita's measurements, we did, however, find oxidative processes on Cu induced by the presence of CO  already at potentials close to the thermodynamic potential for CO oxidation. This result suggests Cu as a very good catalyst for the electrochemical oxidation of \ch{CO}. Cu could thus be an interesting, cheap alternative to the known expensive catalysts for EC CO oxidation e.g.\ in fuel cells, where CO has to be eliminated from the feed to avoid poisoning of especially Platinum catalysts.\cite{Valdes-Lopez2020}

In a joint experimental-theoretical effort, we explore the interaction of Cu with *CO and *OH for alkaline CO oxidation. To this end, we have measured cyclic voltammograms (CVs) with and without CO on well-defined Cu single crystals (SCs). These CVs reveal oxidative processes at the Cu/electrolyte interface in the presence of CO which we interpret as the EC oxidation of CO. We corroborate this hypothesis via \emph{ab initio} simulations demonstrating a feasible reaction mechanism and interface composition at these potentials. We studied Cu SCs with four different surface structures, namely the low-index Cu(100), (110), (111) and (211) orientations. The EC measurements were performed in the respective Cu SC’s fingerprint potential range;  this potential range is positive of the onset of hydrogen evolution and negative of the surface oxidation region. The fingerprint region therefore corresponds to a potential range where the Cu SC is not an oxide or (potentially) a hydride\cite{Huang2018}. The characteristic fingerprint EC response of the first three orientations, which expose large terraces and relatively few steps in absence of any reactive species are known from a previous study (details in Methods section) \cite{Tiwari2020}. Here, we for the first time also present the fingerprint response of (211) (see Figure S1, SI), a prototype of a stepped surface (as further detailed in the SI Sec.\ II). 

\begin{figure}
    \centering
    \includegraphics [width= 0.5\textwidth]{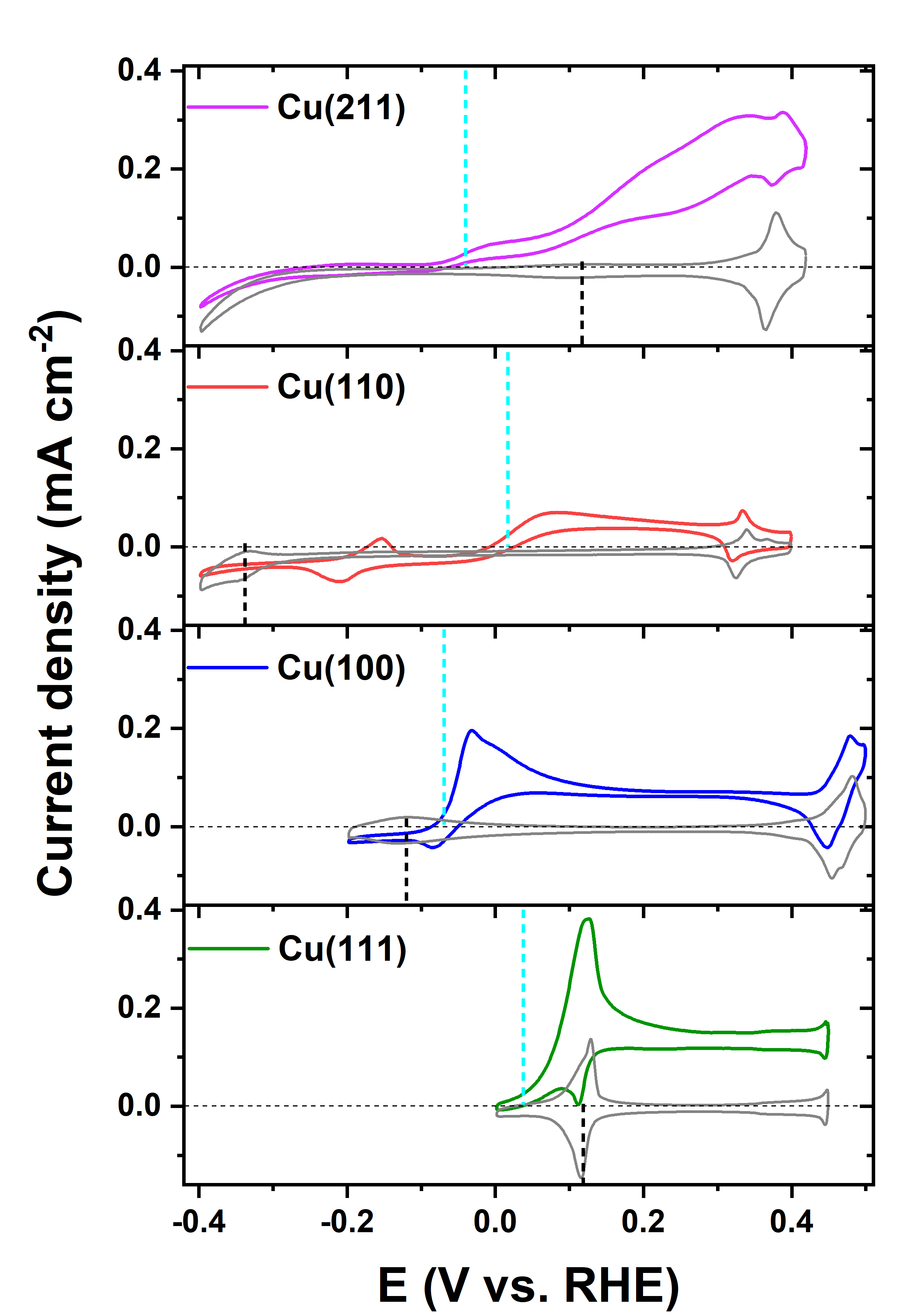}
    \caption{Cyclic voltammograms for Cu(211), (110), (100) and (111) in the respective fingerprint regions measured in blank Ar-saturated (grey trace) and CO-saturated (colored trace) 0.1~M KOH at a scan rate of 50~mV/s. Vertical black dotted lines represent the reversible OH adsorption potential in absence of CO whereas cyan dotted lines mark the onset of CO oxidation.}
    \label{fig:CV}
\end{figure}

\cref{fig:CV} shows CVs of the four Cu orientations in both blank (Ar-) and CO-saturated electrolyte. These measurements show that (a) all four facets have a distinct fingerprint response in blank electrolyte (grey traces), and (b) the fingerprint CVs show clear oxidative features due the presence of CO in the electrolyte (colored traces). In the presence of CO, all facets show increased anodic currents at potentials positive of $-0.1$~V (vs.\ RHE as described further down in  Methods section) which indicates an oxidation process (\emph{onset} indicated by cyan dotted lines). These currents appear slightly positive of both OH-adsorption (known from the fingerprint CVs,\cite{Tiwari2020} \emph{reversible potential} indicated by vertical black dotted lines for better visibility) and the standard equilibrium potential for the CO oxidation reaction $(\ch{CO} + \ch{H2O} \rightarrow \ch{CO2} + \ch{2 H^+} + \ch{2 e^-})$ at $-0.10$~V \cite{Kita1985}. In \cref{tab:potentials}, we list the experimental onset potentials for the reaction, which we define as the potential corresponding to a current density of 25~$\mu$A\,cm$^{-2}$. The oxidative features vanish as soon as CO is purged out of the electrolyte and after which the fingerprint CVs corresponding to an Ar-purged environment are recovered. This reversibility indicates that the surfaces have not undergone any irreversible change in the presence of CO.

\begin{figure}
\centering
\includegraphics [width= 1.0\textwidth]{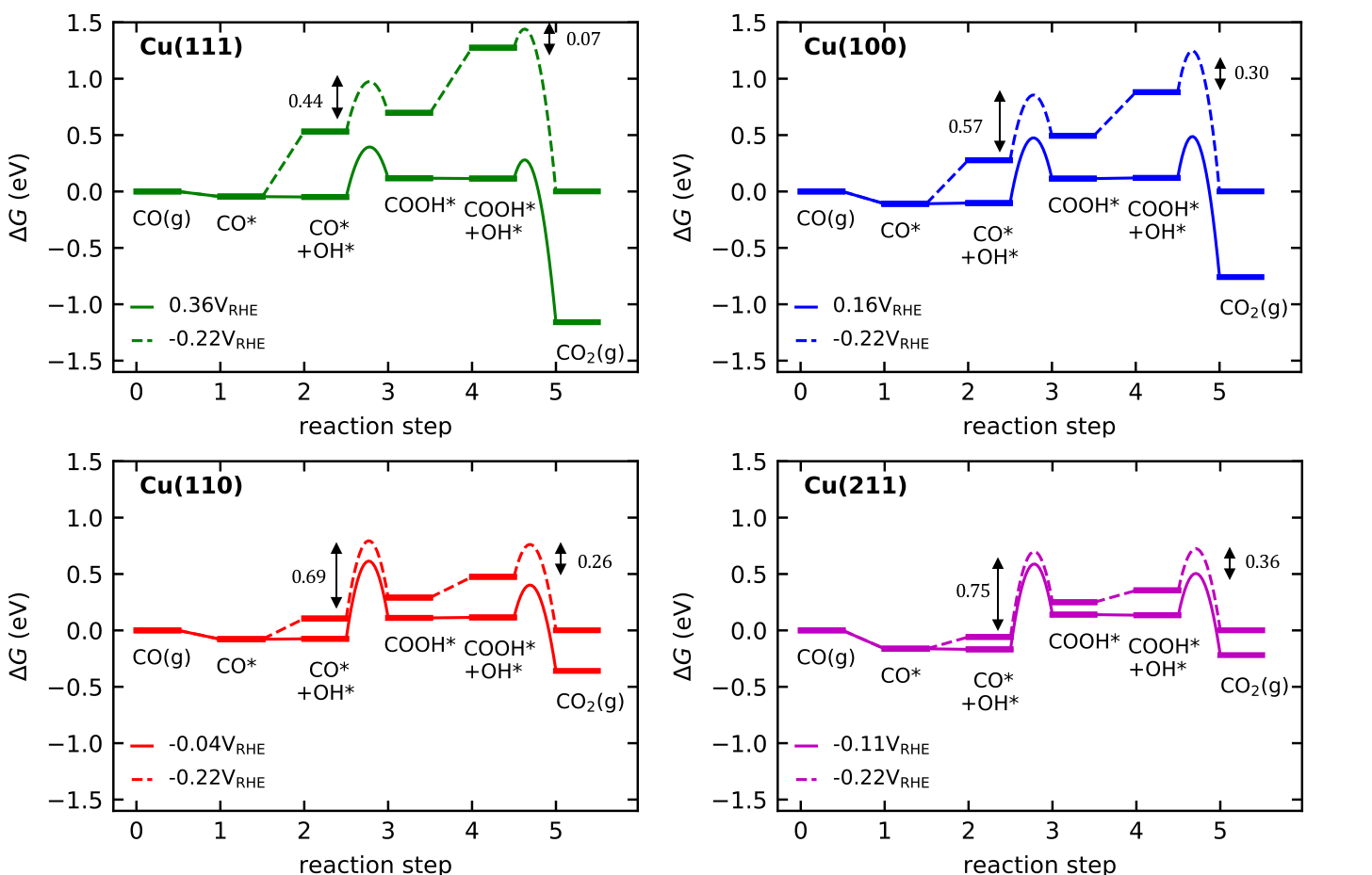}
\caption{Free energy diagram for the suggested CO oxidation reaction mechanism on Cu(111), Cu(100), Cu(110) and Cu(211). The free energy profile is depicted for the onset potential (solid lines) and for the thermodynamic equilibrium potential (dashed lines) as predicted by DFT. The free energies of the chemical (i.e.\ potential-independent) barriers are annotated in eV.}
\label{fig:FEDs}
\end{figure}

To investigate the reaction mechanism and the interplay of *CO and *OH adsorption, we performed \emph{ab initio} simulations based on density functional theory (DFT) of the EC response of the four Cu facets to these adsorbates at anodic potentials. We considered a Langmuir-Hinshelwood mechanism as follows:
\begin{align}
\mathrm{CO(g) +\, ^*} & \rightleftharpoons \mathrm{^*CO} \label{COads} \\
\mathrm{H_2O(l) +\, ^*} & \rightleftharpoons \mathrm{^*OH + H^+ + e^-}  \label{OHads} \\
\mathrm{^*CO +\, ^*OH} & \rightleftharpoons \mathrm{^*COOH} \label{COOHcoupling} \\
\mathrm{^*COOH +\, ^*OH} & \rightleftharpoons \mathrm{H_2O(l) + CO_2(g) + 2^*} \label{COOHOHdissociation}
\end{align}
where * denotes an empty site on the Cu surface.  The corresponding free energy diagrams are shown in  \cref{fig:FEDs}, and we have explicitly considered the association barriers for steps \ref{COOHcoupling} and \ref{COOHOHdissociation}.  
In this mechanism, *OH adsorption (\ref{OHads}) is the sole electrochemical step and the chemical association of *CO and *OH (\ref{COOHcoupling}) is the rate-determining step on all facets considered.   We show in  \cref{fig:FEDs} the energetics both at the theoretically calculated equilibrium potential (-0.22V$_{\text{RHE}}$, dashed lines), as well as at a theoretical onset potential ($E\mathrm{^{Th}_{onset}}$ solid lines), which we define to correspond to the potential where *OH adsorption becomes exergonic.  
At $E\mathrm{^{Th}_{onset}}$, almost all potential-independent reaction thermodynamics and barriers are $< 0.75$~eV and therefore easily surmountable at room temperature \cite{Norskov2014} (c.f.\ \cref{fig:FEDs}). Only on the step-edge of (211), we find the *CO-*OH coupling barrier to be 0.75~eV and therefore a room-temperature activity at the step-edge is equivocal. On the corresponding single crystal, CO oxidation would nevertheless still be feasible on its (111) terrace which could also give rise to the observed current (c.f.\ SI Sec.\ IV). In the proposed reaction mechanism, the Faradaic current arises through continuous replenishment of *OH on all four Cu surfaces which is consumed in the CO oxidation reaction.

In our suggested mechanism, the electrochemical response only depends on *OH adsorption, since this is the only involved electrochemical step. Thus, the facet-dependent adsorption of *OH determines the onset of the EC CO oxidation.  Indeed, as shown in 
\cref{tab:potentials}, the theoretically calculated onset potentials -- solely depending on the interaction-free *OH adsorption energies -- agree within DFT-accuracy with the experimental values. The larger deviation between the theoretical and experimental onset potentials for the large-terrace facets may originate in an underestimated solvation stabilization of *OH through the implicit solvent, as recently demonstrated in Ref.\ \citen{Heenen2020}. As shown in \cref{tab:potentials}, applying an improved solvation correction (c.f.\ SI Sec.\ III) leads to a considerably better agreement.

The experimental onset potentials suggests (100) to be the most active facet towards CO oxidation with an overpotential of 0.03~V which is ca.\ 0.23~V less than Au, previously thought to be the most electroactive element for CO oxidation. \cite{Kita1985} Our proposed mechanism rationalizes this exceptionally low overpotential of CO oxidation on Cu. The potentials at which *OH is predicted to adsorb on the Cu facets and thus CO oxidation sets on (c.f.\ \cref{fig:FEDs}) are very close to the thermodynamic equilibrium potential of CO oxidation (calculated to be -0.22~V which matches the experimentally determined value of -0.1~V within DFT accuracy).
This oxophilicity of Cu is in contrast to Au or Pt, for example, where *OH adsorption occurs significantly more positive \cite{Heenen2020} of the thermodynamic equilibrium potential of CO oxidation, thus leading to higher overpotentials. Due to very high adsorption potentials of *OH in the case of Au, an alternative Eley-Rideal mechanism involving the coupling of *CO with OH$^-$ in solution has been suggested \cite{Edens1996, Blizanac2004a, Duan2018}. 

\begin{table}[hbt!]
\caption{\label{tab:potentials} Experimental onset potentials ($E\mathrm{_{onset}}$), defined to be at 25 $\mu$A\,cm$^{-2}$, as well as the theoretical onset potentials, defined as the onset of *OH adsorption, without ($E\mathrm{_{onset}^{Th}}$) and with ($\tilde E\mathrm{_{onset}^{Th}}$) solvation corrections  (c.f.\ Fig.\ \ref{fig:FEDs}).}
\begin{center}
\begin{tabular}{l|cccc}
\hline
 & $E\mathrm{_{onset}}$ (V) & $E\mathrm{_{onset}^{Th}}$ (V) & $\tilde E\mathrm{_{onset}^{Th}}$ (V) \\
\hline
Cu(111) & 0.04 & 0.36 & 0.18 \\
Cu(100) &  -0.07 & 0.16 & 0.14 \\
Cu(110) &  0.02 & -0.04 & 0.02 \\
Cu(211) &  -0.05 & -0.11 & -0.03 \\
\hline
\end{tabular}
\end{center}
\end{table}

In general, \cref{fig:CV} shows that, on the facets exposing large terraces (100 and 111) the onsets of CO oxidation (cyan dotted lines) and reversible *OH adsorption (black dotted lines) in absence of CO essentially coincide to within 0.1~V. In contrast, on the step rich facets (211 and 110) CO oxidation occurs positive of the onset of *OH adsorption on the step-edges in an Ar atmosphere which we find at -0.35~V for (110) (c.f.\ \cref{fig:CV}) and theoretically estimate to be $\leq$ -0.4~V for (211) (see SI Sec.\ II).

\begin{figure}
\centering
\includegraphics [width= 0.5\textwidth]{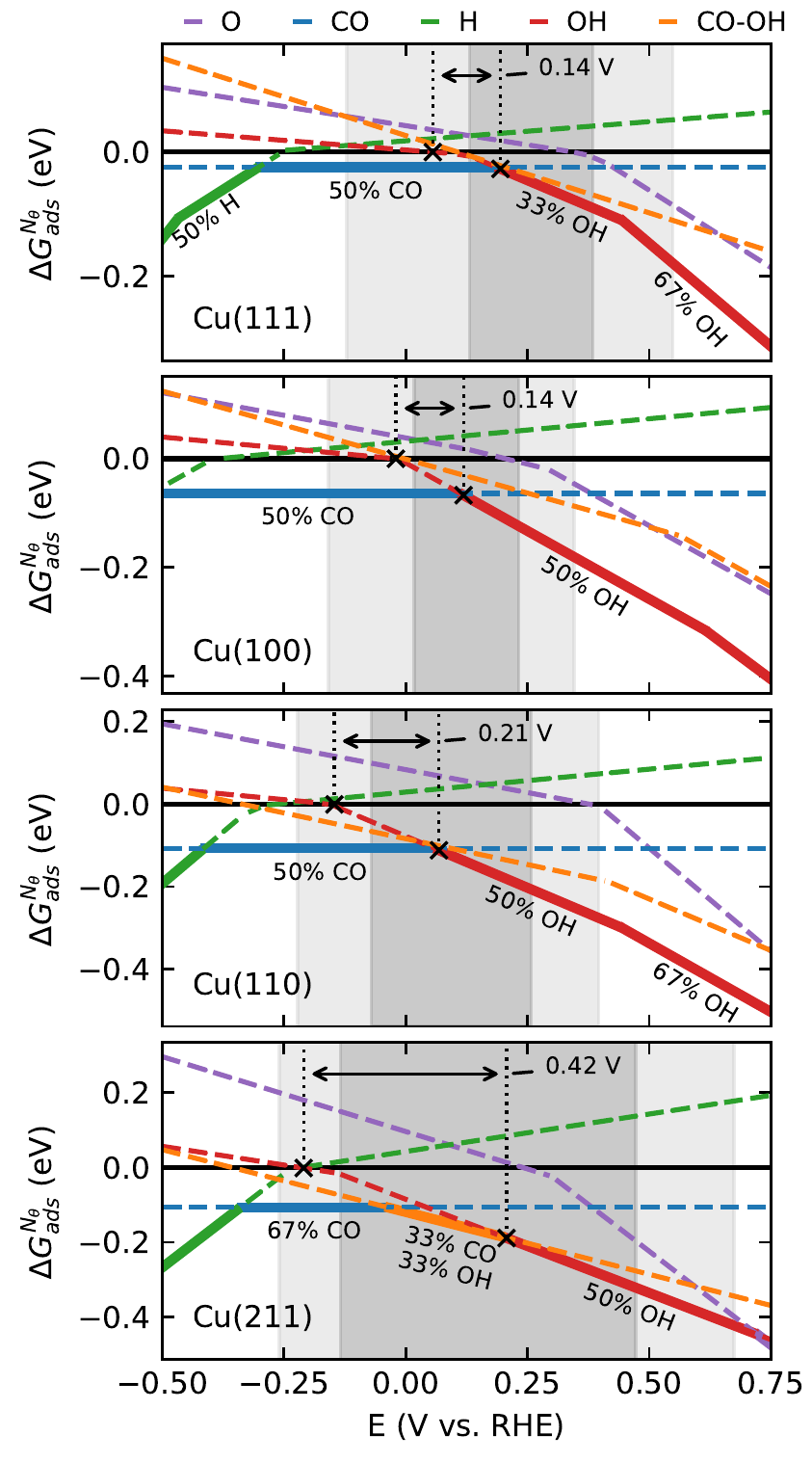}
\caption{Pourbaix analysis of the adsorbates H* (green), *OH (red), *CO (blue), O*
(purple), and co-adsorbed *CO/*OH (orange) on Cu(111), Cu(100),
Cu(110), and Cu(211). The black solid line represents the clean Cu
surface. The dashed lines represent the most stable configuration of each adsorbate (over varying coverage) and the labeled solid lines the overall most stable configuration over the potential range E. The energy $\Delta G_{ads}^{N_\theta}$ is the adsorption energy normalized to the coverage (i.e.\ per site). The shaded area corresponds to the potential range in which the coverage of *CO and *OH are $\geq$ 5\% (dark grey) and $\geq$ 1\% (light grey) as estimated from a Boltzmann distribution (see text). The dotted lines indicate the theoretical interaction-free shift in *OH adsorption potential through the presence of *CO.}
\label{fig:pourbaix}
\end{figure}

We attribute the positive potential shift between CO oxidation and *OH adsorption in the absence of CO to arise from the stronger competitive adsorption of CO vs.\ OH on the steps. This effect is apparent from the theoretical surface Pourbaix diagrams, shown in \cref{fig:pourbaix}. From the configurations sampled, the onset of *OH adsorption in the presence of *CO is shifted +0.14 V on the terraces, and +0.21/+0.42~V on the (110)/(211) steps. This competitive effect can also be seen through the additional/shifted peak at -0.15~V in a CO atmosphere on 110. Following our DFT simulations we, however, do not assign this feature to the *OH adsorption on the 110 step-edge but to a different site, e.g.\ on the 1$\times$2 reconstructed (110) facet \cite{Tiwari2020}. \cref{fig:CV} does not depict an *OH adsorption feature for the step-edge on (211) under Ar atmosphere. Our simulations suggest *OH adsorption on the (211) step edge to occur $\leq$ -0.4~V vs.\ RHE, and therefore the adsorption peak would be convoluted with HER currents (c.f.\ SI Sec\ 2). Nevertheless, without the presence of CO, we still observe a broad 0.12~V peak which corresponds to *OH adsorption on the corresponding (111) terrace of the (211) single crystal (see Fig. S1, SI) \cite{Tiwari2020}. In presence of CO, no *OH adsorption features can be identified on (211). We believe these features, which are subjected to adsorption competition in presence of CO and would thus be found at more positive potentials, to be hidden within the CO oxidation onset.

We note two observations on the potential dependence of the Faradaic currents (c.f.\ \cref{fig:CV}): 
\begin{enumerate}
\item On (111), (100) and (110), the Faradaic current is almost constant with respect to potential. This can be rationalized by the presence of a potential independent rate determining step and relatively constant coverages of the adsorbates involved. Alternatively,  a CO diffusion limitation can also lead to a constant current profile.

\item On (211) such a potential independent current is only seen from -0.01 to 0.1~V and a second current with an apparent potential dependence sets in at $\ge 0.1$~V that also levels off as potential increases. We attribute the first Faradaic process to oxidation of *CO adsorbates on the (111) terrace portion on (211), where the chemical *CO-*OH coupling is more feasible. The onset of this current is more positive than for the (111) single crystal, which suggests that the *OH is initially bound to the stronger binding (211) step edge (see SI Sec.\ IV for a rationalized pathway). We speculate that the second Faradaic process may also arise from the oxidation of CO on (211), but at a different active site (i.e.\ the step edge where the direct *OH/*CO coupling is not very facile). Due to its apparent potential dependence, this alternative process may arise from an electrochemical rate determining step (i.e.\ *CO + OH$^-$ $\rightarrow$ *COOH + $e^-$). 

\end{enumerate}

The proposed mechanism necessitates a finite coverage of both *CO and *OH on the Cu surfaces for the chemical reaction steps to take place. This condition must therefore be fulfilled at all potentials where CO oxidation is active (i.e.\ $\approx 0.0\ldots 0.4$~V). 
We investigate the composition of surface coverage by a Pourbaix analysis \cite{Hansen2008, Bukas2018}. For this analysis, we sample the adsorbates O*, H*, *OH, and *CO on the individual Cu facets. To introduce a coverage dependence, we include each adsorbate in a series of different supercells and further conduct a configurational sampling for *OH and *CO and mixed *OH/*CO (which are dominating at the potentials of interest) in confined supercells enabling (mixed) coverages from 25--100\,\% (see details SI Sec.\ III).
We approximate the potential dependent adsorbate coverages via the sampled configurations for each surface using a Boltzmann distribution. From the most stable free energies $G_i(E)$ of each surface configuration $i$ with an adsorbate coverage $\theta_i^{ads}$, we compute the total coverage $\theta_{tot}^{ads}$ at the potential $E$ for each adsorbate $ads$: 
\begin{align*}
\theta_{tot}^{ads}(E) = \sum_i^N \theta_i^{ads} \exp\left(\frac{G_i(E)}{\mathrm{k_B}T}\right) / Z(E) \\
\mathrm{with} \; Z(E) = \sum_i^N \exp\left(\frac{G_i(E)}{\mathrm{k_B}T}\right)
\end{align*}
where $N$ is the number of sampled compositions and $Z(E)$ the potential dependent partition function. While providing indicative potential dependent surface coverages, we note that the computed partition function is not exhaustive since our sampling is conducted in a limited configurational space. Additionally, the specific interaction with water may affect the considered adsorbate-adsorbate interactions \cite{Kristoffersen2018} and could lead to slight deviation from an idealized behaviour assumed in our analysis. 
The shaded areas in \cref{fig:pourbaix} highlight the potentials at which the coverages of *CO ($\theta_i^{*\mathrm{CO}}$) and *OH ($\theta_i^{*\mathrm{OH}}$) are $\geq$ 5\% or $\geq$ 1\%, respectively, according to our Pourbaix analysis. It can be clearly seen, that finite coverages of both *CO and *OH are present at the potentials where CO oxidation occurs. We note that, while *OH and *CO compete for adsorption sites -- as evident by the shifted *OH adsorption peaks in the CV spectra (c.f.\ \cref{fig:CV}) -- we nevertheless find them to moderately co-stabilize in mixed coverage regimes. 

In regard of the suggested Langmuir-Hinshelwood mechanism, the following observations can finally be summarized in its support:
\begin{enumerate}
    \item coexistence of *OH and *CO at the relevant potential ranges
    \item low *OH-*CO coupling barriers
    \item the CO oxidation onset potential coincides with the *OH-adsorption (as theoretically predicted and experimentally confirmed)
\end{enumerate}
Further, a Langmuir-Hinshelwood mechanism and a chemical reaction as a rate determining step is also in line with previous reports of CO oxidation on Pt(111) \cite{Lebedeva2002}. We, nevertheless, note that an alternative Eley-Rideal mechanism involving a direct oxidation of adsorbed *CO through H$_2$O cannot be definitely excluded.

In summary, we experimentally and theoretically show that Cu is active for the EC oxidation of CO. We find that the corresponding overpotentials are surprisingly low compared to Au, which has been considered as the most electroactive catalyst so far under alkaline conditions. Furthermore, our theoretical calculations suggest that Cu's superior overpotential can be attributed to the facile adsorption of *OH on the catalyst surface as well as the surmountable *CO-*OH coupling barriers. We suggest that Cu is more active than Au or Pt because the latter adsorb *OH at much more positive potentials. Additionally, a structural dependence towards CO oxidation was observed between the four measured orientations, which arises from the facet specificity of *OH and *CO co-adsorption. Our findings highlight the electrocatalytic capability of Cu for CO oxidation due to its fine balance of adsorption energies for the involved adsorbates.
Further, this work suggests Cu as a promising low-cost candidate for EC purification of CO from gas feed in fuel cells. 

\vspace{0.5cm}
\noindent
\textbf{{\large{Methods}}} 

Measurements were performed in a conventional EC setup (described elsewhere \cite{Tiwari2019, Tiwari2020}) in either Ar- or CO-saturated 0.1 M KOH electrolyte and all potentials are referred against reversible hydrogen electrode (RHE). The SCs were cleaned by electropolishing \cite{Tiwari2020} in a 66\% \ch{H3PO4} (85\% EMSURE, Merck) at 2.0 V for 60 s followed by rinsing with Millipore water (18.2~M$\Omega$cm) prior to recording the CVs. The cleanliness of the SCs was confirmed on a regular basis with \emph{ex situ} XPS after the measurements and no traces of contamination especially metals were found. The potential ranges for the CV measurements for each fingerprint region in absence of CO in the electrolyte were purposely chosen in order to avoid the extremes of hydrogen evolution and surface oxidation to allow focus on the structure and potential dependence of the characteristic OH adsorption/desorption features.\cite{Tiwari2020} All the measurements have been performed at least thrice and on two
independent sets of Cu SCs to ensure reproducibility.

Density functional theory (DFT) based on the BEEF-vdW \cite{Wellendorff2012} exchange-correlation functional as implemented in Quantum-Espresso \cite{Giannozzi2009} was used. The electrochemical environment was mimicked through an implicit solvent \cite{Andreussi2012, environ} and electrochemical reactions were referenced via the computational hydrogen-electrode \cite{Norskov2004}. DFT energies were corrected for CO gas-phase reaction thermodynamics \cite{Christensen2015}. Atomic structures, geometry optimizations and nudged-elastic-band (NEB) calculations \cite{Henkelman2000_1, Henkelman2000_2} were handled via ASE\cite{Larsen2017, Torres2019, Koistinen2017, Delrio2019, Boes2019} (see further details in the SI Sec.\ III).

\vspace{0.5cm}
\noindent
\textbf{{\large{Note added in proof}}} 

We were made aware of a parallel publication on CO oxidation over Cu(111) in alkaline electrolyte \cite{Auer2020} that appeared during the review of this paper. Our experimental results are consistent with their work and their electrochemical infrared spectroscopy data supports our mechanistic description. However, we have different theoretical interpretations of the active site and Cu's inherent activity.

\begin{suppinfo}
Separate representation of the Ar and CO saturated CV spectra; detailed discussion of *OH adsorption peaks on Cu(211); details of DFT calculations; detailed depiction of reaction barriers for CO oxidation reactions.
\end{suppinfo}

\begin{acknowledgement}
This work was supported by research grant 9455 from VILLUM FONDEN and the European Union's Horizon 2020 research and innovation programme under the Marie Sklodowska-Curie grant agreement no. 713683.
\end{acknowledgement}

\bibliography{main_theory, references}

\end{document}


\section{Separate CV representation in Ar and CO }\label{exp}

\begin{figure}[H]
    \centering
    \includegraphics [width= 1\textwidth]{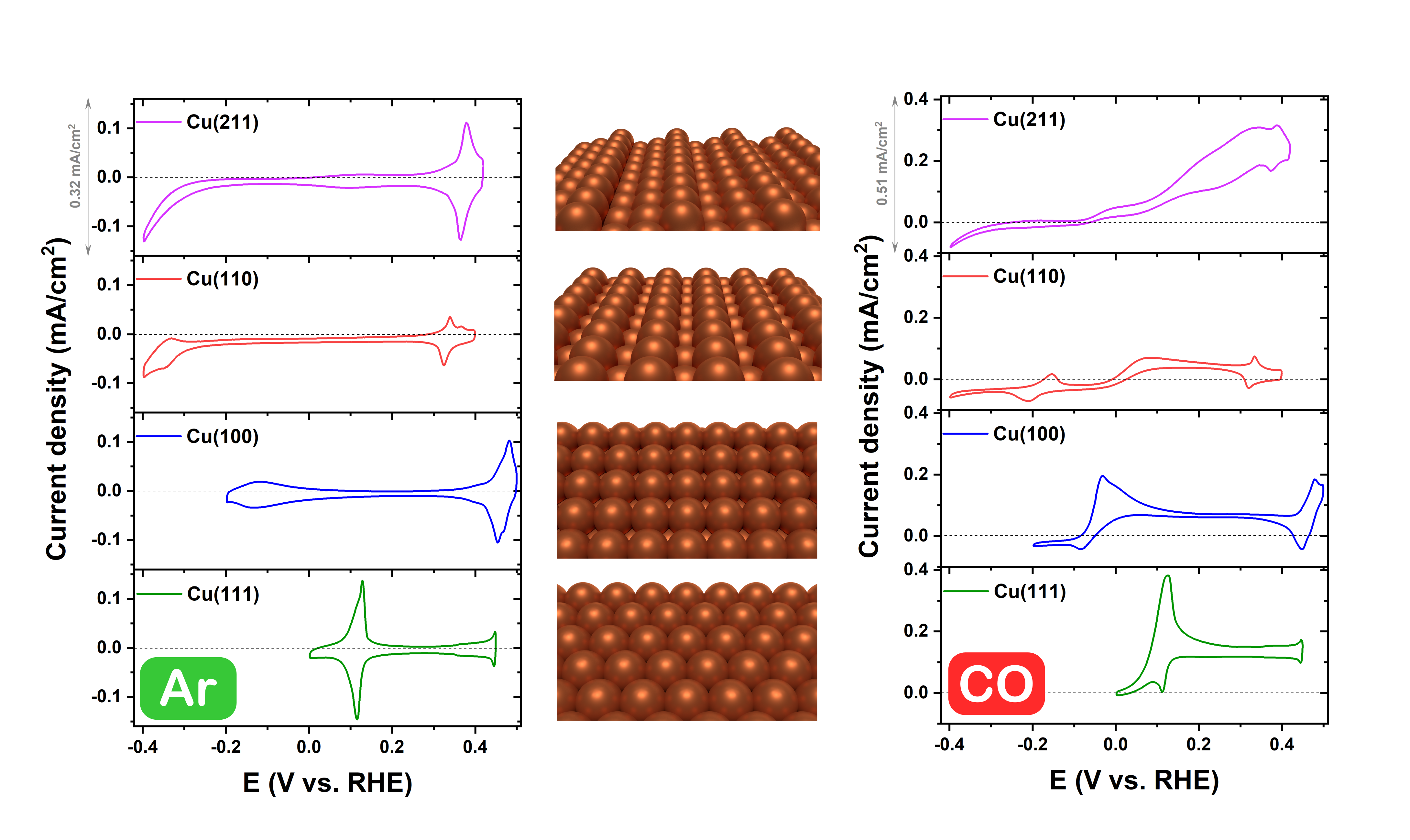}
    \caption{Cyclic voltammograms and corresponding atomistic representation (central panel) for Cu(211), Cu(110), Cu(100) and Cu(111) in their respective fingerprint regions measured using a conventional EC setup in blank Ar-saturated (left panel) and CO-saturated (right panel) 0.1~M KOH electrolyte at a scan rate of 50~mV/s.
    The CVs presented in this figure are the same as in Figure 1, but presented individually for better visibility.}
    \label{fig:CV_SI}
\end{figure}

\section{*OH adsorption peaks at the electrolyte/Cu(211) interface}\label{exp}

For Cu(211) DFT predicts a lowest *OH adsorption free energy on the bridge site on the step edge at: $\Delta G_{ads}$= -0.175 eV (see Table \ref{tab:Engads}) which -- excluding interactions, configurational entropy and solvation -- corresponds to an adsorption at -0.175 V$_{\mathrm{RHE}}$. Relating this adsorption energy to Cu(110), which itself has a DFT predicted adsorption free energy of $\Delta G_{ads}$=-0.089 eV \cite{Tiwari2020}, we would expect an idealized *OH adsorption potential shift of -0.086 V$_{\mathrm{RHE}}$ for Cu(211) in comparison to Cu(110). Since the *OH adsorption peak on Cu(110) is just recognizable within the onset of the HER region\cite{Tiwari2020} we subsequently expect the *OH adsorption peak on the Cu(211) step edge deep in the HER region and thus not detectable; assuming a similar onset of HER on Cu(211) and Cu(110).

\begin{figure*}[!hbtp]
\includegraphics[width=0.5\textwidth]{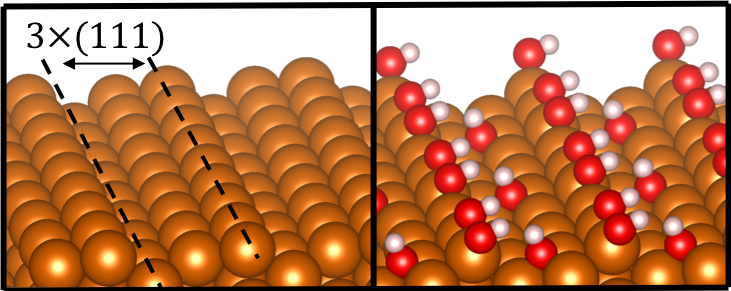}
\caption{Atomistic presentation of the Cu(211) facet without coverage of adsorbates (left) and with *OH decorated bridge-step sites and *OH coverage on the (111) portion of the Cu(211) facet (right). The dashed lines indicate the (111) portion of the Cu(211) facet.}
\label{fig:Cu211atoms}
\end{figure*}

The visible *OH peak in the Cu(211) CV (around 0.1 V) likely corresponds to the (111) facet portion of the (211) crystal as the measured potential is the same as in the CV of the (111) as shown in Fig.\ \ref{fig:CV_SI}. The integrated charge corresponds to only 23\,\% of the peak found on the Cu(111) single crystal. The available (111) facet on the Cu(211) SC corresponds to $\approx$ 66.8 \% of the geometric area (compare to Fig.\ \ref{fig:Cu211atoms}). When *OH adsorbs on the (111) portion of the Cu(211) SC, the edges are already *OH decorated and thereby occupy and repel, obstructing adsorption on some available sites on the (111) portion. As suggested in Fig. \ref{fig:Cu211atoms} the free sites on the (111) portion correspond to only 1/3 yielding a total of 22.3\, \% of the (111) facet. This lies closely to the experimental value.

\section{Computational Details for atomistic simulations}\label{sec:1}

We performed density functional theory (DFT) calculations based on the BEEF-vdW \cite{Wellendorff2012} exchange-correlation functional,  a plane wave basis set and ultrasoft pseudopotentials as implemented in the Quantum ESPRESSO code \cite{Giannozzi2009}. The plane-wave and density cutoffs were 500 and 5000\,eV, respectively, and a Fermi-smearing width of 0.1 eV was applied. We converged the electronic structure until a total energy difference of $10^{-5}$\,eV. 

Solvation effects were included via the self-consistent continuum solvation (SCCS) model \cite{Andreussi2012} as implemented in the Environ package \cite{environ}. We reparameterized the implicit solvent model to reproduce the experimental potential of zero charge of Cu(111), Cu(100), and Cu(110) and a capacitance of 20\,$\mathrm{\mu F}$ as suggested by H{\"o}rmann \textit{et al.} \cite{Hormann2019}. The resulting electronic density cutoffs were determined to be $\rho_{\mathrm{min}} = 0.0001$ and $\rho_{\mathrm{max}} = 0.0145$. To reproduce implicit solvation energies within $\leq$ 0.05\,eV of the original SCCS parameterization, we adjust the effective surface tension term to $(\alpha + \gamma) = 19.9$\,dyn/cm. We stress, that the calculations presented in this work are near-identical to the calculations conducted with the original SCCS parameterization. Our reparameterization would only affect calculations including implicit charging which is mainly critical when calculating charge transfer reactions \cite{Nattino2019, Gauthier2019_unified} and was not resorted to in this work. The solvation stabilization of adsorbates on metal surfaces -- especially those who undergo hydrogen binding with water -- is underestimated through implicit solvents, as recently demonstrated \cite{Heenen2020}. For the *OH adsorbate we therefor apply a suggested \textit{posteriori} correction to estimate the error in the solvation stabilization. The correction is based on a scaling with the *OH binding energy ($\Delta E\mathrm{_{ads}^{vac, *OH}}$) following the relation: $\Delta E\mathrm{_{solv}^{*OH}} = -0.22 \mathrm{eV} - 0.44 \cdot \Delta E\mathrm{_{ads}^{vac, *OH}} \mathrm{eV}$ \cite{Heenen2020}. The corrected values are separately presented in Tab.\ I in the main text.

We used symmetric slab models to simulate the adsorbates on the Cu single-crystal surface. For the Cu(111), Cu(100), Cu(110) and the reconstructed Cu(211) facet, supercells of the sizes 3$\times$4, 4$\times$3, 2$\times$3, and 1$\times$4 with vacuum spacing of 12 \AA \,  on each side were used, to avoid interactions of adsorbates across periodic images. We converged the slab thicknesses to 4, 5, 9, and 4 atom layers with the central 2, 3, 3, and 2 layers constrained, respectively. The Brillouin zone was sampled via a 4$\times$3$\times$1, 3$\times$4$\times$1, 3$\times$4$\times$1, and 4$\times$3$\times$1 Monkhorst-Pack grid \cite{Monkhorst1976}, respectively. For nudged-elastic-band (NEB) calculations \cite{Henkelman2000_1, Henkelman2000_2}, non-symmetric slab models were used to accommodate the higher computational demand and difficulty in convergence. Additionally, for the NEB calculations of the *COOH + *OH $\rightarrow$ H$_2$O(l) + CO$_2$(g) barrier on the Cu(110) facet an extended 2$\times$4 (Brillouin sampling 3$\times$3$\times$1) was employed to avoid interaction across periodic boundaries along the step edge.
For the surface Pourbaix analysis, we considered the adsorbates O*, H*, *OH, and *CO. To introduce a coverage dependence, we sampled each adsorbate on its most stable site on a series of different supercells down to a size of 1$\times$1 (containing only 1 adsorption site). Further, we conducted a configurational sampling for *OH and *CO and mixed *OH/*CO at their respective most stable sites. In this sampling, we considered all symmetrically inequivalent configurations in a 2$\times$2 supercell for Cu(111), Cu(100), and Cu(110) and in a 1$\times$3 supercell for Cu(211) enabling mixed coverages between 25\,\% and 100\,\%. Overall, the sampled surface configurations for the surface Pourbaix analysis included $\approx$ 180 structures.

For the handling of atomistic structures, geometry optimizations, vibration calulations and NEB calculations the Atomic Simulation Environment (ASE) package \cite{Larsen2017} was employed.  Geometry optimizations were converged until a force of 0.05\,eV/\AA . For all adsorbates, the most stable adsorption sites (see Tab.\ \ref{tab:Engads}) were identified through sampling of all symmetrically inequivalent adsorption sites as identified through the package CatKit \cite{Boes2019}. Barrier calculations were conducted using the NEB method and handled by AIDNEB\cite{Torres2019, Koistinen2017, Delrio2019}. The barriers were computed with a minimal accuracy of the surrogate model of 25\,meV and until a convergence of the forces on the climbing image of 25-50\,meV. All transition states were confirmed to have a singular imaginary frequency. Free energies were obtained following the ideal gas law for gas phase species and the harmonic oscillator model for adsorbates, respectively \cite{Norskov2014}. All energies were referenced to gas-phase H$_2$ (1 atm), CO (1 atm) and H$_2$O at the vapor pressure of liquid water (0.035 atm) \cite{CRCHandbook}. To mitigate systematic DFT errors, we apply the corrections suggested by Christensen \textit{et al.}\cite{Christensen2015} of 0.15\,eV per C=O double bond and 0.1\,eV for the H$_2$(g) reference.

\begin{table}[hbt!]
\caption{\label{tab:Engads} DFT adsorption energies for the *OH, *CO, and *COOH adsorbates on the most stable sites of the investigated Cu surfaces. Given are the adsorption sites, electronic adsorption energies $\Delta E$ and the adsorption free energies $\Delta G$ including the corrections by Christensen \textit{et al.}\cite{Christensen2015} (see text). All energies are referenced to gas-phase H$_2$ (at 1 atm), CO (1 atm) and H$_2$O (at 0.035 atm) and given in eV.}
\begin{center}
\begin{tabular}{l|ccc|ccc|ccc}
\hline
facet &  & *OH & & & *CO & & & *COOH & \\
 & site & $\Delta E$ & $\Delta G$ & site & $\Delta E$ & $\Delta G$ & site & $\Delta E$ & $\Delta G$ \\
\hline
\hline
(111) & bridge / & -0.015 & 0.286 & hollow & -0.545 & -0.288 & hollow & -0.451 & 0.205 \\
 & hollow & 0.018 & 0.327 & & & &  \\
\hline
(100) & bridge & -0.195 & 0.136 & top & -0.647 & -0.350 & bridge & -0.636 & 0.002 \\
\hline
(110) & bridge & -0.389 & -0.066 & top & -0.590 & -0.321 & top & -0.884 & -0.201 \\
\hline
(211) & bridge & -0.471 & -0.145 & top & -0.672 & -0.405 & bridge & -0.923 & -0.242 \\
\hline
\end{tabular}
\end{center}
\end{table}

\clearpage
\section{Reaction barriers for CO oxidation} \label{sec4:barriers}

\begin{figure*}[!hbtp]
\includegraphics[width=0.35\textwidth]{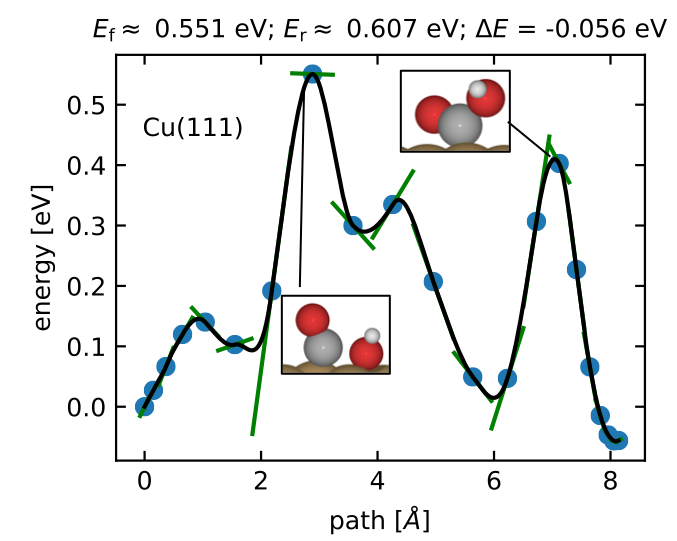} 
\includegraphics[width=0.35\textwidth]{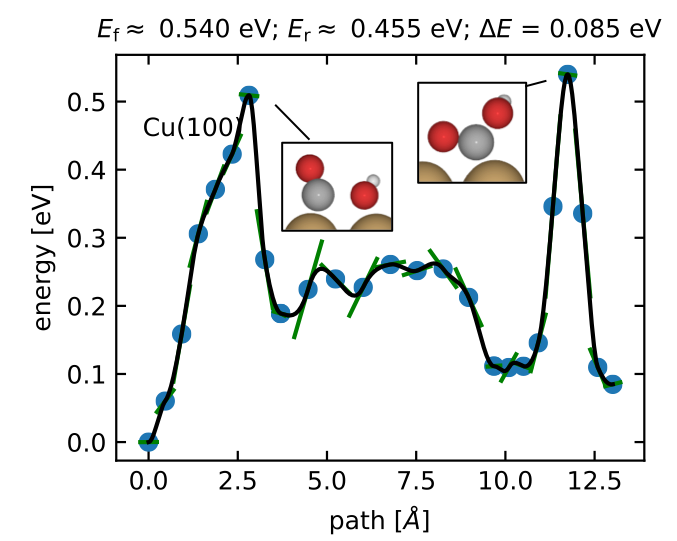} \hfil
\includegraphics[width=0.35\textwidth]{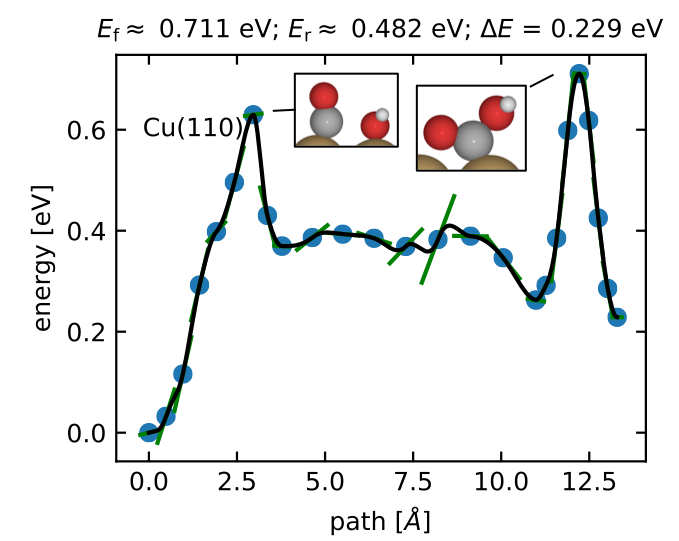} 
\includegraphics[width=0.35\textwidth]{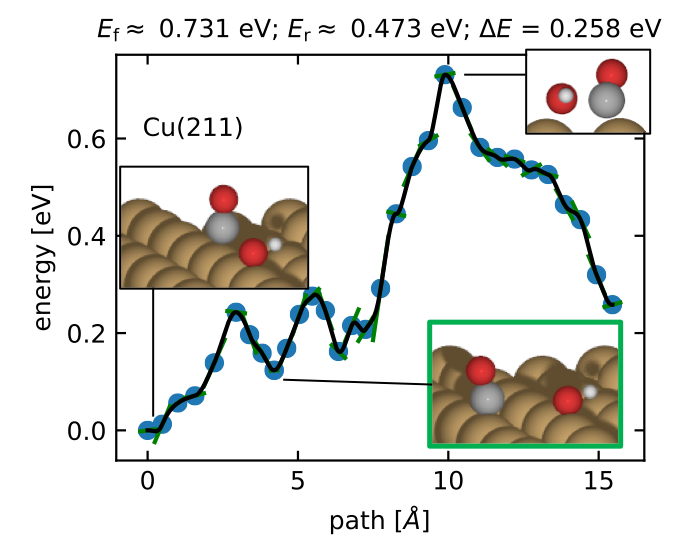}
\caption{Potential energy profile of the converged NEB calculation for the *CO-*OH coupling reaction on Cu(111) (top left), Cu(100) (top right), Cu(110) (bottom left), and Cu(211) (bottom right). The blue points correspond to the individual images and the green lines to their gradient along the minimum energy pathway. The highest energy points correspond to the transition states. Inset pictures depict the atomic configuration of selected images. For Cu(211), the intermediate minimum corresponding to *CO on the (111) portion of the terrace is marked in green (see text). The forward barrier $E_f$, reverse barrier $E_r$, and reaction energy $\Delta E$ are annotated.}
\label{fig:nebCO-OH}
\end{figure*}

\begin{figure*}[!hbtp]
\includegraphics[width=0.35\textwidth]{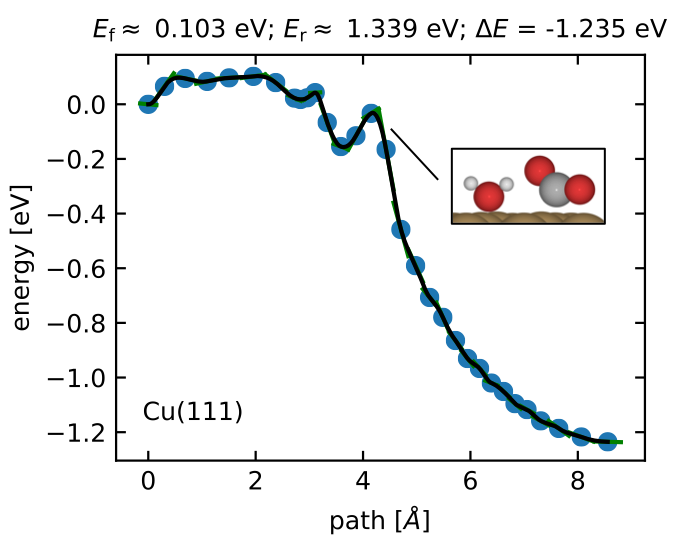} 
\includegraphics[width=0.35\textwidth]{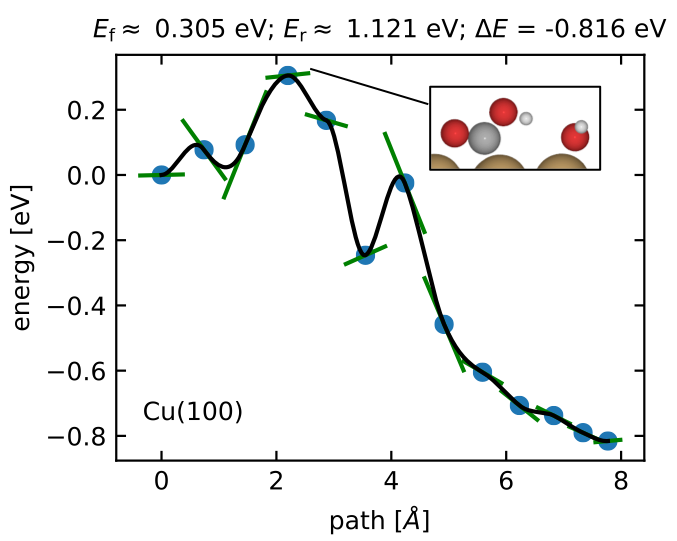} \hfil
\includegraphics[width=0.35\textwidth]{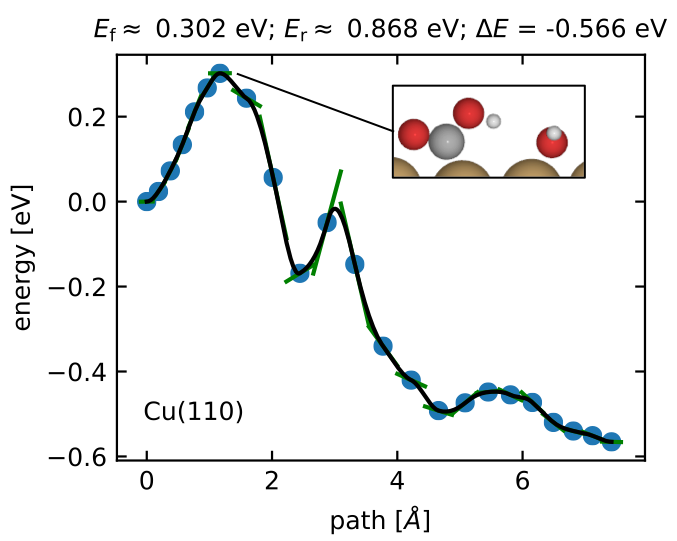} 
\includegraphics[width=0.35\textwidth]{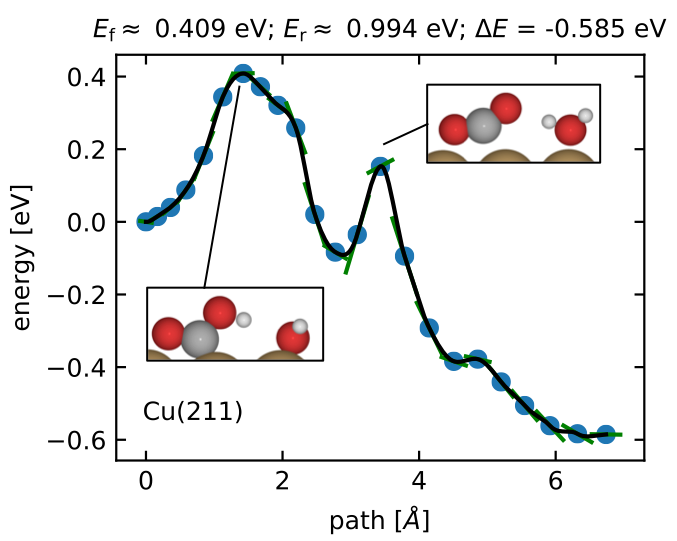}
\caption{Potential energy profile of the converged NEB calculation for the *COOH-*OH deprotonation reaction on Cu(111) (top left), Cu(100) (top right), Cu(110) (bottom left), and Cu(211) (bottom right). The blue points correspond to the individual images and the green lines to their gradient along the minimum energy pathway. The highest energy points correspond to the transition states. Inset pictures depict the atomic configuration of selected images. The forward barrier $E_f$, reverse barrier $E_r$, and reaction energy $\Delta E$ are annotated.}
\label{fig:nebCOOH-OH}
\end{figure*}

The electronic energy profiles of the minimum energy pathways calculated via the NEB method for the *CO-*OH coupling reaction and *COOH-*OH deprotonation reaction are depicted in Fig.\ \ref{fig:nebCO-OH} and \ref{fig:nebCOOH-OH}, respectively. The *CO-*OH coupling reactions appear to accommodate comparably complex pathways including 1-2 local minima and in most cases two transition states. As indicated by the inset atomistic depictions in Fig.\ \ref{fig:nebCO-OH}, the forming *COOH intermediate undergoes a conformational rearrangement (rotation) during the reaction. All included barriers exhibit surmountable potential energy barriers (same for the free energy barriers, see main text). The CO-OH coupling barriers on the step rich facets (110) and (211) appear slightly higher than on the terrace rich facets. Especially on the latter Cu(211) facet, the coupling barrier of $\Delta E^\dagger$ 0.73\,eV ($\approx$ 0.75 eV in free energy) is just surmountable. With a lower coupling barrier of $\approx \Delta E^\dagger$ = 0.6\,eV, the CO-OH coupling reaction on the Cu(211) could alternatively initiate from a *CO adsorbate on the (111) portion of this facet. This intermediate configuration is contained in our computed NEB calculation and marked in green in Fig.\ \ref{fig:nebCO-OH}. Considering a fully *CO/*OH decorated step, additional *CO adsorbates on the (111) portion would be trapped at a higher energy local minimum. These adsorbates could engage in the coupling reaction with a lowered barrier (then starting from the marked configuration) and lead to higher CO oxidation currents.

Compared to the *CO-*OH coupling reaction, the *COOH-*OH deprotonation reaction exhibit low energetic barriers. The contained barriers originate from surface diffusion steps of the *OH adsorbate. In contrast to this, the deprotonation of the *COOH is barrier-less.

The chemical reaction barriers (and the computed minimum energy pathways -- not shown) are relatively insensitive to the application of the implicit solvent model. As shown in Tab.\ \ref{tab:Engbarrier}, a moderate stabilization of $\approx$ 0.2\,eV for the coupling reaction step is seen and a destabilization of $\approx$ 0.2\,eV for the deprotonation step is seen. It follows that the implicit solvent does not influence the qualitative picture of the reaction barriers (or the reaction mechanism).

\begin{table}[hbt!]
\caption{\label{tab:Engbarrier} Electronic energies converged without implicit solvent (``vacuum'') and with implicit solvent (``implicit'') of the reaction barriers *CO-*OH coupling reaction and the *COOH-*OH deprotonation reaction for the different Cu facets.}
\begin{center}
\begin{tabular}{l|cccc}
\hline
$\Delta E ^{\ddagger} _{\mathrm{vacuum}}$ &Cu(111) & Cu(100) & Cu(110) & Cu(211) \\
\hline
*CO-*OH & 0.754 & 0.791 & 0.954 & 0.901 \\
*COOH-*OH & 0.055 & 0.194 & 0.225 & 0.280 \\
\hline
\hline
$\Delta E ^{\ddagger} _{\mathrm{implicit}}$ &Cu(111) & Cu(100) & Cu(110) & Cu(211) \\
\hline
*CO-*OH & 0.551 & 0.540 & 0.711 & 0.731 \\
*COOH-*OH & 0.103 & 0.305 & 0.302 & 0.409 \\
\hline
\end{tabular}
\end{center}
\end{table}

\bibliography{supporting_information}